\documentclass[prb,preprint]{revtex4-1}

\usepackage{amsmath}  
\usepackage{amsfonts} 
\usepackage{graphicx} 

\usepackage[font=small,
   justification=raggedright,
   format=plain]{caption}
\usepackage{subcaption}

\newcommand{\ts}[1]{\textrm{\tiny #1}}

\newcommand{\Ts}{\text{s}}

\begin{document}

\title{Looping Pendulum: Theory, Simulation, and Experiment}

\author{Collin Dannheim}
\email{cdannheim@luc.edu} 
\author{Luke Ignell}
\email{lignell@luc.edu} 
\author{Brendan O'Donnell}
\email{bodonnell3@luc.edu} 
\author{Robert McNees}
\email{rmcnees@luc.edu} 
\author{Constantin Rasinariu}
\email{crasinariu@luc.edu} 
\affiliation{Department of Physics, Loyola University Chicago, Chicago, IL 60626}

\date{\today}

\begin{abstract}
The looping pendulum is a simple physical system consisting of two masses connected by a string that passes over a rod. We derive equations of motion for the looping pendulum using Newtonian mechanics, and show that these equations can be solved numerically to give a good description of the system's dynamics. The numerical solution captures complex aspects of the looping pendulum's behavior, and is in good agreement with the experimental results.
\end{abstract}

\maketitle 

\section{Introduction}

The looping pendulum consists of two different masses, one heavy and one light, connected with
a string that passes over a horizontal rod. One lifts up the heavy mass by pulling down on the light one  at a given angle with the vertical axis. Releasing the light mass, it sweeps around the rod, keeping the heavy mass from falling to the ground.  This mesmerizing physics demo, which we first became aware of through a video on YouTube \cite{YouTube}, appeared as one of the problems in the 2019 International Young Physicists Tournament \cite{IYPT} and was also studied to some degree in a recent paper \cite{Yubo}.\footnote{The model presented in this paper was developed before the publication of reference \cite{Yubo}.} Its simple implementation and spectacular dynamics make it an intriguing problem for physics students and professors alike.

In this paper we study the looping pendulum using Newtonian mechanics, simulate its motion using numerical methods, and compare the theoretical predictions with the experimental results. The experiments closely confirm the numeric simulations and reveal rich dynamics with surprising features. In some cases we notice a hint of sensitivity to small changes in the initial conditions. 

The manuscript is organized as following: in Section \ref{sec:EOM} we derive the equations of motion, and obtain a pair of coupled, second-order, non-linear differential equations. In Section \ref{sec:ExpSetup} we present the experimental setup, and in Section \ref{sec:ExpSim} we compare the experimental results with numerical solutions of the model obtained using Mathematica~\cite{Mathematica}. Finally, in Section \ref{sec:Disc} we review our results, outline plans to improve the apparatus, and consider the possibility that, for a given range of the parameters of the model, the looping pendulum may exhibit chaotic behavior.

\section{Equations of Motion for the Looping Pendulum}
\label{sec:EOM}

\begin{figure}[h!]
  \centering
  	\includegraphics[width=0.66\columnwidth]{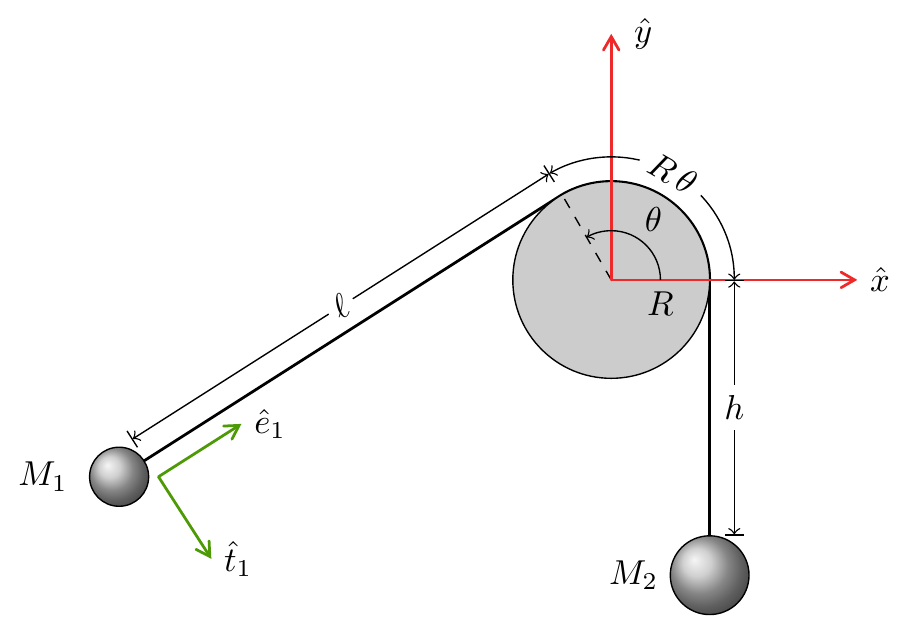}
	\caption{This cross-section view of the looping pendulum shows masses $M_1$ and $M_2$ connected by a string that passes over a rod with radius $R$. The radius of the rod is much smaller than the initial values of $\ell$ or $h$ in our experiments, and is exaggerated here for clarity.}
  \label{PendulumDiagram} 
\end{figure}

The dynamics of the looping pendulum can be described using physics that is typically covered in first-year and second-year undergraduate courses in United States universities. Our setup is shown in Fig.~\ref{PendulumDiagram}.
A heavy mass $M_2$ is suspended vertically by a string that passes over a rigid rod of radius $R$. The distance from $M_2$ to the point where the string makes contact with the rod is $h$. The string winds around the rod for an angle $\theta$, then extends a distance $\ell$ from the rod to a lighter mass $M_1$ which is held at rest. The lighter mass is released, and the system is allowed to move freely. During this motion the quantities $h$, $\theta$, and $\ell$ will change, but the total length of the string remains fixed at $L = \ell + R\,\theta + h$. 

The motion of mass $M_2$ is best described using a fixed coordinate system $\hat{x}$ and $\hat{y}$ with the origin taken at the center of the rod's circular cross section. Mass $M_2$ is assumed to move straight up and down with no horizontal component to its motion. Its position is $\vec{r{}}_2 = R\,\hat{x} - h\,\hat{y}$ and its velocity is $\vec{v}_2 = -\dot{h}\,\hat{y}$, where a dot indicates the derivative with respect to time. Since the length $L$ of the string is fixed, $\dot{L} = \dot{\ell} + R\,\dot{\theta} + \dot{h} = 0$ and the velocity and acceleration of $M_2$ can be written as:
\begin{gather}\label{eq:v2a2}
	\vec{v{}}_2 = (\dot{\ell} + R\,\dot{\theta})\,\hat{y}~, \qquad \vec{a{}}_{2} = (\ddot{\ell} + R\,\ddot{\theta})\,\hat{y} ~.
\end{gather}

The lighter mass $M_1$ follows a more complicated path. Its position is given by
\begin{gather}
	\vec{r{}}_{1} = \big( R\,\cos\theta - \ell\,\sin\theta \big)\,\hat{x} + \big( R\,\sin\theta + \ell\,\cos\theta\big)\,\hat{y} ~.
\end{gather}
However, it is convenient to describe its motion using a co-moving system of coordinates that point parallel and perpendicular to the string connecting $M_1$ to the rod. We define the orthonormal unit vectors $\hat{e}_{1}$ and $\hat{t}_{1}$ (shown in Fig.~\ref{PendulumDiagram}) as
\begin{gather}\label{eq:eunit}
	\hat{e}_{1} = \sin\theta\,\hat{x} - \cos\theta\,\hat{y} ~,\\ \label{eq:tunit}
	\hat{t}_{1} = -\cos\theta\,\hat{x} - \sin\theta\,\hat{y} ~.
\end{gather}
Then the position of $M_1$ can be written as
\begin{gather}
	\vec{r{}}_{1} = -\ell\,\hat{e}_{1} - R\,\hat{t}_{1} ~.
\end{gather}
Because these new unit vectors depend on $\theta$, they will change as the string loops around the rod. Their first and second derivatives with respect to time are
\begin{gather}
	\dot{\hat{e}}_{1} = -\dot{\theta}\,\hat{t}_{1}\,, \qquad \dot{\hat{t}}_{1} = \dot{\theta}\,\hat{e}_{1}\,, \\
	\ddot{\hat{e}}_{1} = -\ddot{\theta}\,\hat{t}_{1} - \dot{\theta}^{\,2} \hat{e}_{1} \,, \qquad 
	\ddot{\hat{t}}_{1} = \ddot{\theta}\,\hat{e}_{1} - \dot{\theta}^{\,2} \hat{t}_{1} ~.
\end{gather}
Using these results, we obtain for the velocity and acceleration of $M_1$ 
\begin{gather} \label{eq:v1}
	\vec{v{}}_{1} = - \big(\dot{\ell} + R\,\dot{\theta}\big)\hat{e}_{1} + \ell\,\dot{\theta}\,\hat{t}_{1}\,, \\ \label{eq:a1}
	\vec{a{}}_{1} = \big( \ell\,\dot{\theta}^{\,2} - \ddot{\ell} - R\,\ddot{\theta}\big) \hat{e}_{1} 
		+ \big( 2\,\dot{\ell}\,\dot{\theta} + \ell\,\ddot{\theta} + R\,\dot{\theta}^{\,2} \big) \hat{t}_{1}~.
\end{gather}
As a check, notice that the $\hat{e}_{1}$ component of $\vec{v}_{1}$ -- the rate at which the string is moving onto or off of the rod -- is minus the rate at which $M_2$ is moving downwards or upwards in Eq.~\eqref{eq:v2a2}. 

Now that we have expressions for the accelerations of each mass, we need to identify the forces acting on them. For mass $M_2$ there is an upward force due to a tension $\vec{F}_{T,2}$, a downward force $\vec{F{}}_{g,2}$ due to gravity, and a small but non-negligible amount of air resistance $\vec{F{}}_{\text{air},2}$ in the direction opposite to $\vec{v{}}_{2}$. These forces are
\begin{gather}
	\vec{F{}}_{T,2} = T_2\,\hat{y}\,, \qquad \vec{F{}}_{g,2} = -M_2\,g\,\hat{y}\,, \qquad \vec{F{}}_{\text{air},2} = -\gamma_2\,|\vec{v{}}_2|\,\vec{v{}}_{2} ~,
\end{gather}
where the coefficient $\gamma_2$, discussed in more detail in Section \ref{sec:ExpSim}, encodes the dependence of $\vec{F{}}_{\text{air},2}$ on the drag coefficient of $M_2$ and the area it presents in a plane perpendicular to $\vec{v{}}_{2}$. Using Eq.~\eqref{eq:v2a2}, the net force on mass $M_2$ is
\begin{gather}
	\vec{F}_{\text{net},2} = \big( T_2 - M_2\,g - \gamma_2\,|\dot{\ell} + R\,\dot{\theta}|\,(\dot{\ell} + R\,\dot{\theta})\big)\,\hat{y} ~.
\end{gather}
Notice that the force due to air resistance changes sign based on the direction of $\vec{v}_{2}$. 
\begin{figure}[h]
  \centering
    \includegraphics[width=0.66\columnwidth]{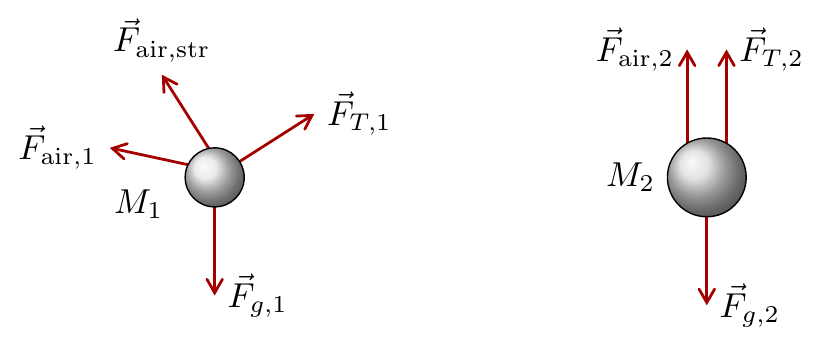}
	\caption{ \label{FreeBodyDiagrams}
The forces acting on $M_1$ and $M_2$ include gravity, tension in the string, and air resistance. 
}
 \end{figure}
For mass $M_1$ there are four forces: a tension $\vec{F}_{T,1}$ pulling along the direction $\hat{e}_{1}$ of the string, gravity pulling downwards, air resistance in the direction opposite to $\vec{v}_{1}$, and an effective force associated with air resistance experienced by the string. The first three forces have the form
\begin{gather}
	\vec{F{}}_{T,1} = T_1\,\hat{e}_{1}~, \qquad \vec{F{}}_{g,1} = -M_{1}\,g\,\hat{y}~, \qquad \vec{F{}}_{\text{air},1} = -\gamma_1\,|\vec{v{}}_{1}|\,\vec{v{}}_{1} ~.
\end{gather}
The final force acting on $M_1$ comes from drag on the string connecting it to the rod. Because drag forces are proportional to area, and the diameter of the string's cross section is much smaller than its length, we approximate this force as being due entirely to the area of the string in a plane perpendicular to $\hat{t}_{1}$ and ignore its motion in the $\hat{e}_{1}$ direction. Therefore,
\begin{gather}
	\vec{F}_{\text{air},\text{str}} \simeq - \gamma_{\text{str}}\,|\vec{v}_{1}\cdot \hat{t}_{1}|\,(\vec{v}_{1}\cdot \hat{t}_{1})\,\hat{t}_{1} ~.
\end{gather}
The area of the string factoring into $\gamma_\text{str}$ is proportional to its length $\ell$, which changes as the masses move. To account for this, we write $\gamma_\text{str} = \ell\,\alpha_\text{str}$. Using Eqs.~\eqref{eq:eunit}-\eqref{eq:tunit} and Eq.~\eqref{eq:v1}, the net force on $M_1$ is then
\begin{equation}
\begin{split}
	\vec{F}_{\text{net},1} = \big( T_1 + M_1\,g\,\cos\theta + \gamma_1\,|\vec{v}_{1}|\,(\dot{\ell}+R\,\dot{\theta})\big)\,\hat{e}_{1} \\ 
	+ \big(M_1\,g\,\sin\theta - \gamma_1\,|\vec{v}_{1}|\,\ell\,\dot{\theta} - \alpha_{\text{str}}\,|\ell\,\dot{\theta}|\,\ell^{2}\,\dot{\theta}\big) \, \hat{t}_{1} ~.
\end{split}	
\end{equation}
The coefficients $\gamma_{1}$ and $\alpha_{\text{str}}$ characterizing the air-resistance forces will be discussed in more detail in Section \ref{sec:ExpSim}. 

The tensions $T_1$ and $T_2$ will differ because of friction between the string and the rod. 
\begin{figure}[h!]
  \centering
  	\includegraphics[width=0.66\columnwidth]{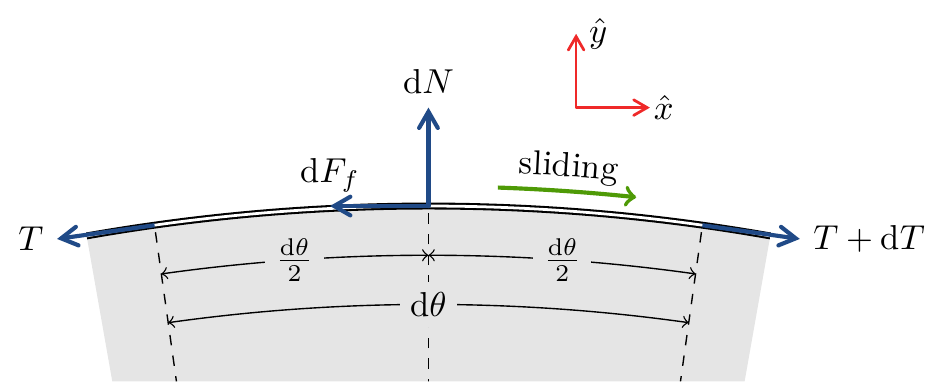}
	\caption{\label{Tensions}
Friction between the string and the rod determines the difference between the tensions $T_1$ and $T_2$.}
  \end{figure}
Figure \ref{Tensions} shows an infinitesimal segment of string covering an angle $\text{d}\theta$ along the rod. We assume here that the string is sliding across the surface of the rod in the clockwise direction. This produces a small amount of friction $\text{d}F_{f} = \mu\,\text{d}N$ in the counterclockwise direction, where $\mu$ is the coefficient of friction between the string and the rod and $\text{d}N$ is the normal force acting on the string segment. Since $\text{d}\theta \ll 1$, we use $\cos \frac{\text{d}\theta}{2} \simeq 1$ and $\sin \frac{\text{d}\theta}{2} \simeq \frac{\text{d}\theta}{2}$ to write the net force on the segment of string as
\begin{gather}
	\vec{F}_{\text{net}} = (\text{d}T - \mu\,\text{d}N) \,\hat{x} + (\text{d}N - T\,\text{d}\theta)\,\hat{y} ~.
\end{gather}
Requiring the net force to vanish gives 
\begin{gather}
	\text{d}T = T\,\mu\,\text{d}\theta ~.
\end{gather}
Integrating $\text{d}T$ over the interval between the two points where the string makes contact with the rod, as illustrated in Fig.~\ref{PendulumDiagram}, gives the relationship between $T_1$ and $T_2$:
\begin{gather}\label{eq:T2T1}
	T_2 = T_1\,e^{-\text{sign}(v_2)\, \mu\,\theta} ~.
\end{gather}
The factor of $-\text{sign}(v_2)$ in the exponent accounts for the fact that the string may slide along the rod in either direction, depending on whether $M_2$ is moving downwards or upwards. The clockwise sliding shown in Fig.~\ref{Tensions} corresponds to $M_2$ moving downwards, in which case $\text{sign}(v_2) < 0$ and $T_2 > T_1$.

We are now in position to write the equations of motion for the looping pendulum. For mass $M_1$, the $\hat{e}_1$ and $\hat{t}_1$ components of Newton's second law give 
\begin{align}\label{eq:M1e1EOM}
	M_1\,\big(\ell\,\dot{\theta}^{\,2} - \ddot{\ell} - R\,\ddot{\theta}\big) ={} &  T_1 + M_1\,g\,\cos\theta +\gamma_{1}\sqrt{\big(\dot{\ell}+R\,\dot{\theta}\big)^{2} + \big(\ell\,\dot{\theta}\big)^{2}}\,\big(\dot{\ell}+R\,\dot{\theta}\big)
	~, \\
	\label{eq:M1t1EOM}
	M_1\,\big( 2\,\dot{\ell}\,\dot{\theta} + \ell\,\ddot{\theta} + R\,\dot{\theta}^{\,2} \big) ={} & M_1\,g\,\sin\theta - \gamma_1\,\sqrt{\big(\dot{\ell}+R\,\dot{\theta}\big)^{2} + \big(\ell\,\dot{\theta}\big)^{2}}\,\ell\dot{\theta} - \alpha_{\text{str}}\,|\ell\,\dot{\theta}|\,\ell^{\,2}\dot{\theta} ~.
\end{align}
For mass $M_2$ the equation of motion is
\begin{gather}
	M_2\,\big( \ddot{\ell} + R\,\ddot{\theta}\big) = T_2 - M_2\,g - \gamma_2\,|\dot{\ell} + R\,\dot{\theta}|\,\big(\dot{\ell} + R\,\dot{\theta}\big) ~.
\end{gather}
Solving Eq.~\eqref{eq:M1e1EOM} for $T_1$ and using Eq.~\eqref{eq:T2T1} leads to the following pair of coupled, second-order, non-linear differential equations for $\ell(t)$ and $\theta(t)$:
\begin{align} 
  \label{eq:Eqn1} %
		0 = {} &  2\,\dot{\ell}\,\dot{\theta} + \ell\,\ddot{\theta} + R\,\dot{\theta}^{\,2} - g\,\sin\theta + \frac{\gamma_1}{M_1}\,\sqrt{\big(\dot{\ell}+R\,\dot{\theta}\big)^{2} + \big(\ell\,\dot{\theta}\big)^{2}}\,\ell\,\dot{\theta} + \alpha_{\text{str}}\,|\ell\,\dot{\theta}|\,\ell^{\,2}\dot{\theta}\,, \\ 
  \label{eq:Eqn2} %
		0 = {} & \ddot{\ell} + R\,\ddot{\theta} + g + \frac{\gamma_2}{M_2}\,|\dot{\ell} +R\,\dot{\theta}|\,\big(\dot{\ell} + R\,\dot{\theta}\big) 
		 \\ \nonumber
		 & - e^{-\text{sign}(v_{2}) \mu\,\theta} \left( \frac{M_1}{M_2}\,\big(\ell\,\dot{\theta}^{\,2} - \ddot{\ell} - R\,\ddot{\theta} - g\,\cos\theta\big) - \frac{\gamma_{1}}{M_2}\sqrt{\big(\dot{\ell}+R\,\dot{\theta}\big)^{2} + \big(\ell\,\dot{\theta}\big)^{2}}\,\big(\dot{\ell}+R\,\dot{\theta}\big) \right) ~.
\end{align}
These equations cannot be solved in closed form, so in this paper we use Mathematica to solve them numerically. The numerical solution is compared with our experimental results in Section \ref{sec:ExpSim}.

The phenomenon that first drew our attention to the looping pendulum is the tendency of mass $M_2$ to fall some distance from its initial position and then stop moving. In some cases its motion will stop temporarily before falling again, or it may even briefly move upwards. But once $M_2$ stops completely -- a behavior we refer to as \emph{halting} -- the subsequent motion of $M_1$ simplifies somewhat. First, the shape of its trajectory (assuming it is moving fast enough to maintain tension in the string) is entirely determined by the action of the string wrapping around the rod. Second, if air resistance is ignored then the time dependence of $\ell$ and $\theta$ are well-approximated by a simple function. These observations, which aren't essential in what follows, are discussed in more detail in Appendix \ref{app:SubsequentMotion}.

\section{Experimental Setup}
\label{sec:ExpSetup}

To test the model derived in Section \ref{sec:EOM} we constructed several looping pendulums and recorded the motion of masses $M_1$ and $M_2$ for a range of initial conditions. The $x$ and $y$ positions of each mass were then extracted from a slow-motion video using the Tracker Video Analysis and Modeling Tool~\cite{Tracker}.

\begin{figure}[h!]
  \centering
	\includegraphics[width=0.66\columnwidth]{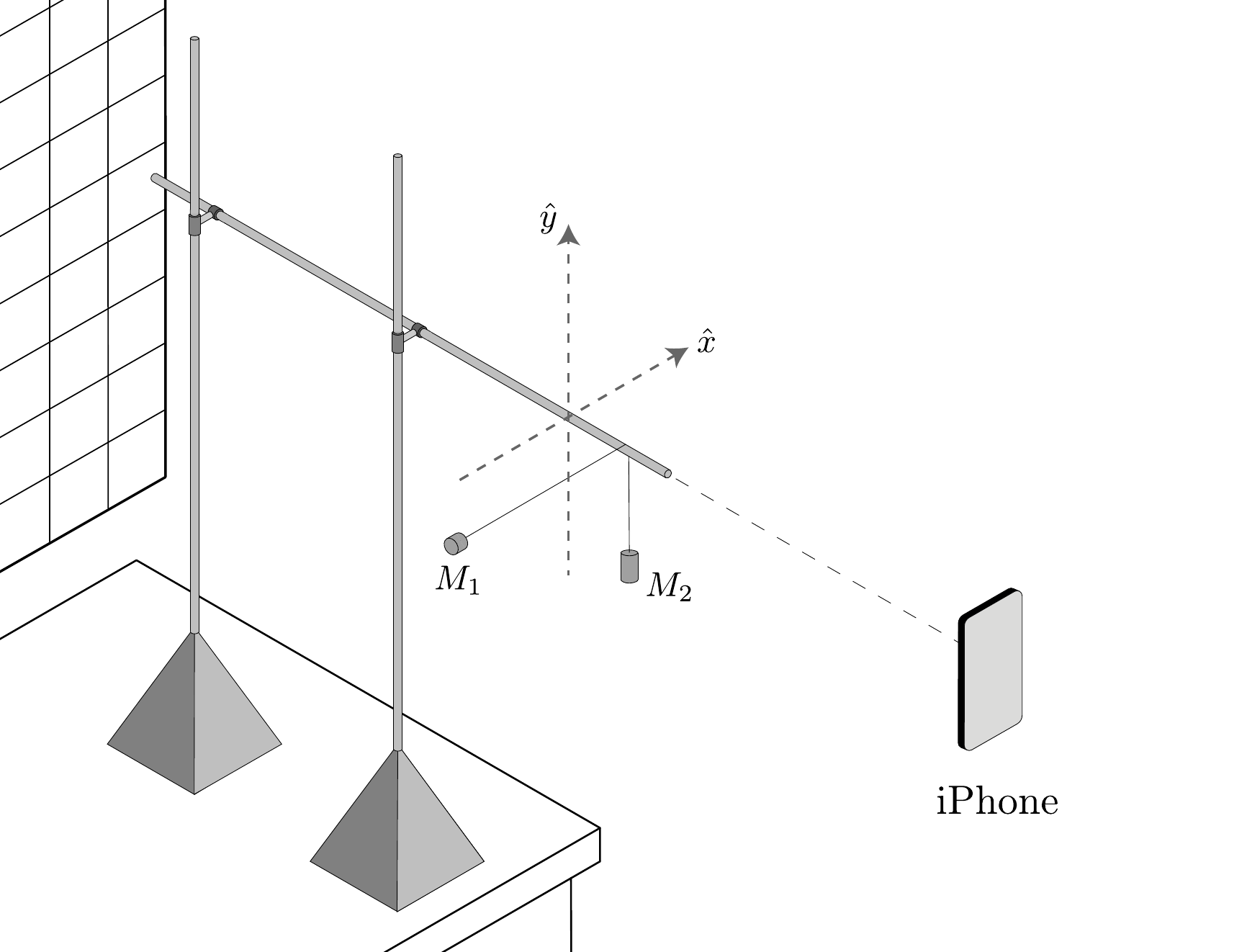}
	\caption{  \label{Apparatus} The arrangement of the rods, masses, string, and camera is shown in this schematic of the apparatus used in our experiments.}
\end{figure}

A schematic of our apparatus is shown in Fig.~\ref{Apparatus}. The frame consisted of three aluminum rods. Two vertical rods were mounted on lab stands, with clamps connected to each rod supporting a third, horizontal rod. The third rod, which was aligned using a level, appears in cross section in Fig.~\ref{PendulumDiagram}. The two masses were connected by a string which was placed over this rod. The lighter mass was then released from rest at different initial angles. The subsequent motion of the masses was recorded in a 240 fps video using an iPhone placed in front of the apparatus. A large white panel with a grid drawn on it was mounted behind the apparatus to improve alignment of the camera and visibility of the masses.

The materials used in the apparatus were selected because they were on hand at the beginning of the project. With the exception of a set of cylindrical masses used in later trials, all materials were taken from one of the undergraduate project rooms in the Physics Department at Loyola University Chicago. Multiple pendulums were built following the setup described above, but there was some variation in materials. A number of different aluminum rods were used for the horizontal rod. While all the rods had a length of approximately $1\,\text{m}$ and a diameter between $1.2$-$1.3\,\text{cm}$, rough spots and variations in surface finish gave different values for the coefficient of friction $\mu$. The string connecting the masses was taken from a spool of generic kite string and was not selected for any particular properties other than its negligible mass. The masses themselves initially comprised bundles of washers with individual masses in the range of $7.5$-$7.7\,\text{g}$. The irregular shape of the bundles made it difficult to model air resistance so they were eventually replaced with a set of cylindrical masses. Later versions of the apparatus also corrected problems like alignment of the camera, using a leveled tripod to mount the iPhone rather than a standard lab clamp. 

There were several dimensionless parameters that could be varied when running the experiment: the ratio $M_2/M_1$ of the masses; various ratios of the lengths $L$, $\ell$, and $R$; and the initial angle $\theta_i$ for the point of contact between the string and the rod. We observed the behavior of the masses for several values of $M_2/M_1$, multiple values of the initial angle $\theta_i$ for each value of the mass ratio, and fixed \footnote{Fixed values within small variations associated with positioning the string, attaching the masses, and using different rods} values of $L$, $\ell$, and $R$.

A run of the experiment began by selecting two masses that gave a specific value of $M_2 / M_1$. The masses were connected by a string, a spot was marked on the string where it would make contact with the rod, and the total length of the string $L$ and the length $\ell$ from the point of contact to $M_1$ were measured. The mass $M_1$ was then positioned at different initial angles, held at rest but with tension in the string, and released. The participant recording the run would scrutinize the video for signs that the masses had made contact with each other or with the string. If there was any indication of contact the run would be repeated. This happened frequently, so getting a ``clean'' run with no collisions typically required several attempts. This process was then repeated for a new initial angle, with multiple initial angles tested for each mass ratio. 

Once a run was successfully completed, the 240 fps video was trimmed to begin at the frame where the masses were released and loaded into the Tracker Video Analysis and Modeling Tool. The length scale was set by identifying the distance between two points in the first frame with one of the measured lengths (typically $\ell$), and then verified by comparing the computed and measured values of the other lengths ($L$, $R$, and $h$). The initial angle that the string connected to $M_1$ made with the horizontal direction was measured and converted into the initial angle $\theta$ shown in Fig.~\ref{PendulumDiagram}. Then the positions of each mass were recorded frame-by-frame until mass $M_1$ reached the rod -- typically resulting in between 200 and 400 data points. The positions were marked on screen at the points where the string attached to the masses. Once all the positions had been marked, the data was exported to CSV files containing the $(t, x, y)$ data points for each mass. The data was then imported into Mathematica and compared with the numerical solution of our model.

There were aspects of the apparatus which could not be incorporated into the model of Section \ref{sec:EOM}. For example, the model assumes that the trajectory of the masses lies in a plane. However, it was difficult to realize this in practice. When the masses experienced coplanar motion they often collided with each other or with the string. To avoid this, the trajectory of mass $M_1$ had to have a small component in the direction parallel to the rod, which is not included in our simulation. Another problem arose from the string looping over itself as it wound around the rod. This changed the value of the coefficient of friction in a way that we could not account for in the model. Fortunately, these effects could be minimized: a careful release of the mass reduced motion parallel to the rod, and a trial could be repeated if the video showed the string looping over itself and subsequently slipping. More significant sources of error which can be addressed in future experiments with different materials or equipment are discussed in Section \ref{sec:Disc}.

\section{Experimental Results and Comparison with Simulation}
\label{sec:ExpSim}

To compare our experimental results to the model of Section \ref{sec:EOM} we solve Eqs.~\eqref{eq:Eqn1}-\eqref{eq:Eqn2} numerically. The equations depend on several parameters which must be measured or estimated. The quantities $R$, $L$, $\ell$, and $\theta_i$ were measured directly and/or extracted from video using the Tracker Video Analysis and Modeling Tool, while the coefficient of friction $\mu$ was determined using a procedure described below. The drag coefficients could not be measured and were instead estimated based on the shape and dimensions of the masses and string.

\begin{figure}[h]
  \centering
  	\includegraphics[width=0.25\columnwidth]{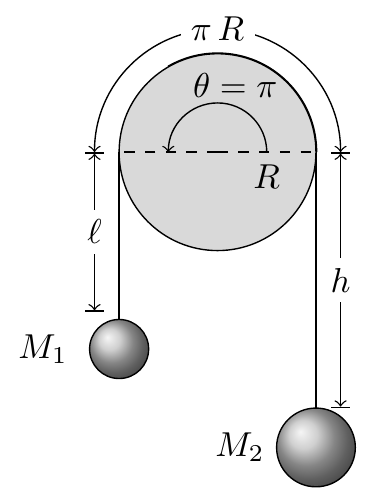} 
	\caption{  \label{PendulumFriction}
 The masses are suspended vertically and released from rest. Mass $M_2$ moves downwards with a constant acceleration (ignoring air resistance) that depends on $g$, $\mu$, and the ratio $M_2 / M_1$.}
\end{figure}

\begin{figure}[h]
	\includegraphics[width=0.75\columnwidth]{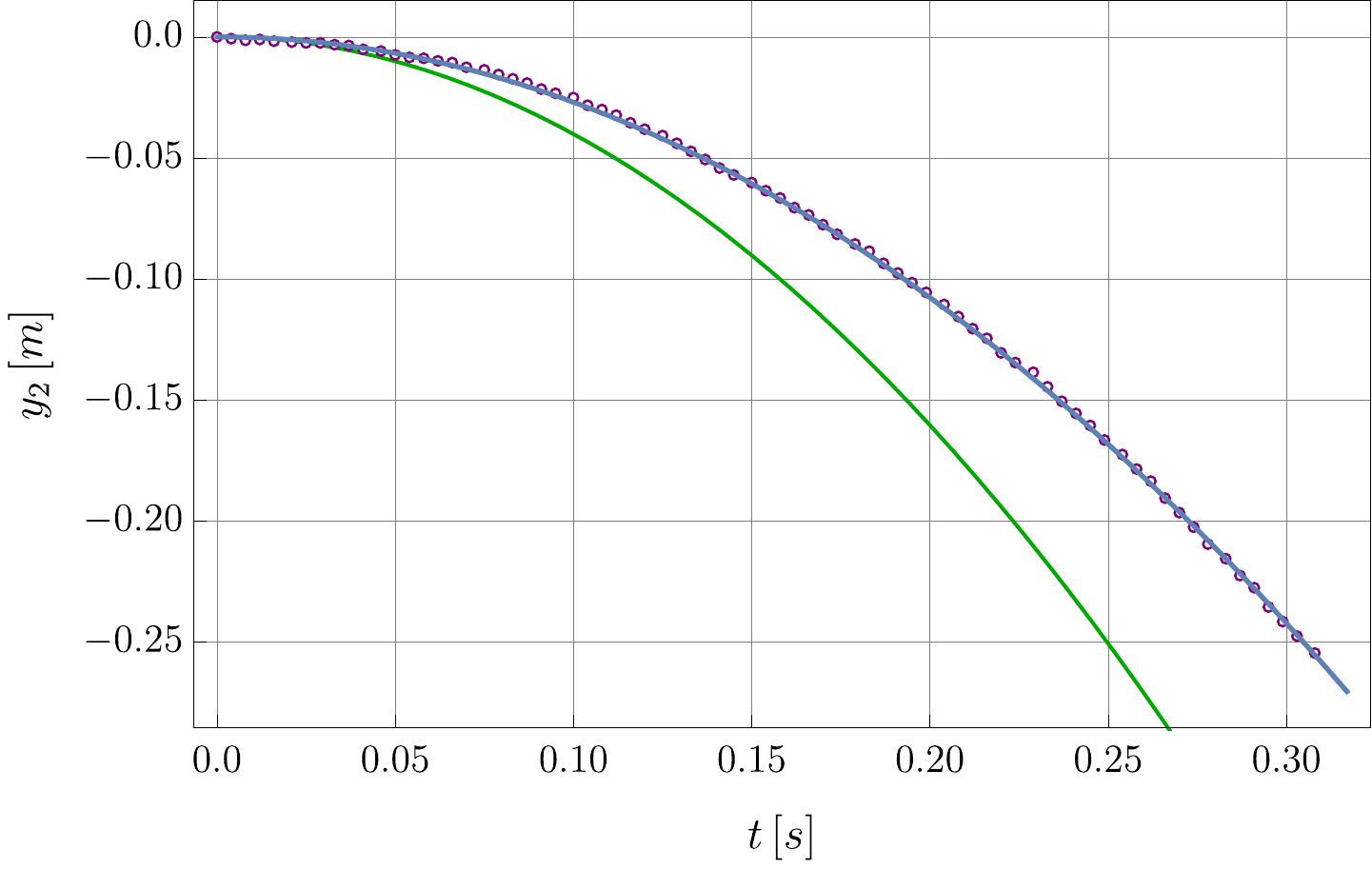}
	\caption{\label{Friction10to1} For the case $M_2/M_1 = 10$, the green curve shows the trajectory with $\mu=0$, the purple circles show the positions extracted from video of the falling mass, and the blue curve represents the best-fit value $\mu=0.34$ for Eq.~\eqref{FallingVertically}.} 
\end{figure}

The coefficient of friction $\mu$ between the string and rod was determined by tracking the motion of masses $M_1$ and $M_2$ after they were suspended vertically and released from rest. This arrangement is shown in Fig.~\ref{PendulumFriction}. In that case the masses move up or down with constant acceleration. The motion of the masses was recorded, the position of $M_2$ as a function of time was extracted from the video using Tracker, and $\mu$ was determined by fitting this data to the trajectory
\begin{gather}\label{FallingVertically}
	y_{2}(t) = y_{2}(0) - \frac{1}{2}\,g\,\frac{M_2 - M_1\,e^{\mu\pi}}{M_2 + M_1\,e^{\mu\pi}}\,t^2 ~.
\end{gather}
An example of a best-fit for the case $M_2/M_1 = 10$, yielding the value $\mu=0.34$, is shown in Fig.~\ref{Friction10to1}.

Air resistance of the masses and string was more difficult to quantify. In Eqs.~\eqref{eq:Eqn1}-\eqref{eq:Eqn2} the three coefficients $\gamma_{1}$, $\gamma_{2}$, and $\alpha_{\text{str}} = \gamma_{\text{str}}/\ell(t)$ determine the magnitude of the drag forces. These coefficients are generally expected to take the form
\begin{gather}
	\gamma = \frac{1}{2}\,C\,\rho_{\text{air}}\,A ~,
\end{gather}
where $C$ is a drag coefficient of order 1, $\rho_{\text{air}} \simeq 1.2\,\text{kg}/\text{m}^{3}$ is the density of air, and $A$ is the area that the object presents in a plane perpendicular to its direction of motion. In early runs of the experiment the masses comprised bundles of washers. The irregular shape of the bundles made estimates of $A$ difficult, especially as $M_1$ changed orientation during its motion, so the washers were later replaced with a set of cylindrical masses. Assuming a drag coefficient $C$ of order 1, and treating the area $A$ as that of a rectangle with dimensions given by the height and diameter of the cylinder, the coefficients $\gamma_1$ and $\gamma_2$ were estimated to both be of order $10^{-3}\,\text{kg}/\text{m}$. For the string connecting $M_1$ to the rod, the area in the plane perpendicular to $\hat{t}_{1}$ is equal to its length $\ell(t)$ times its diameter. The string used in the experiment has a diameter of approximately $2\,\text{mm}$, so the ratio $\alpha_{\text{str}} = \gamma_{\text{str}} / \ell(t)$ is estimated to be $\alpha_{\text{str}} \simeq 10^{-3} \,\text{kg}/\text{m}^{2}$.

\begin{figure}[htb]
     \centering
     \begin{subfigure}[b]{0.48\textwidth}
         \centering
         \includegraphics[width=\textwidth]{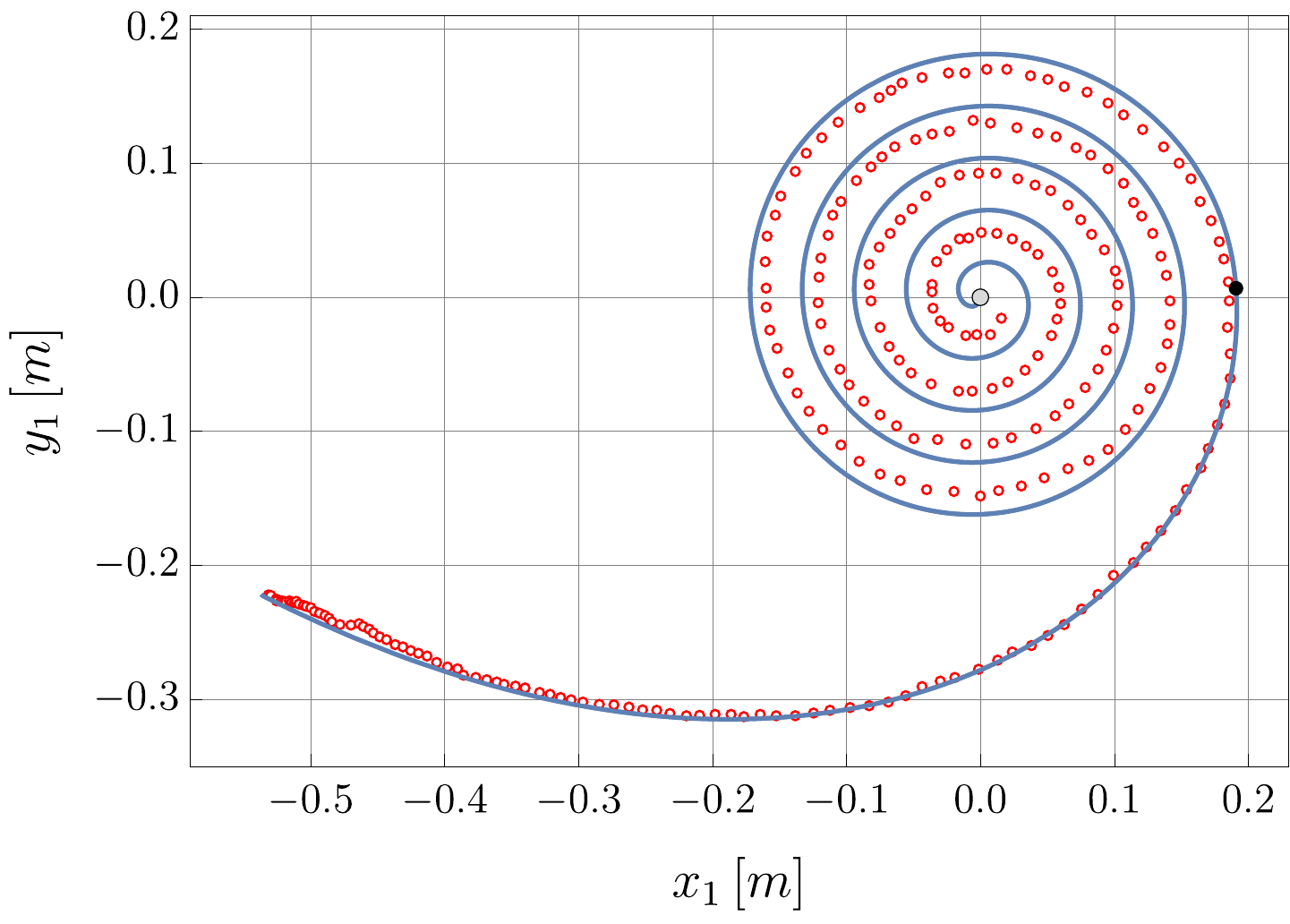} 
         \caption{}
         \label{fig:M1Trajectory}
     \end{subfigure}
     \hfill
     \begin{subfigure}[b]{0.48\textwidth}
         \centering
         \includegraphics[width=\textwidth]{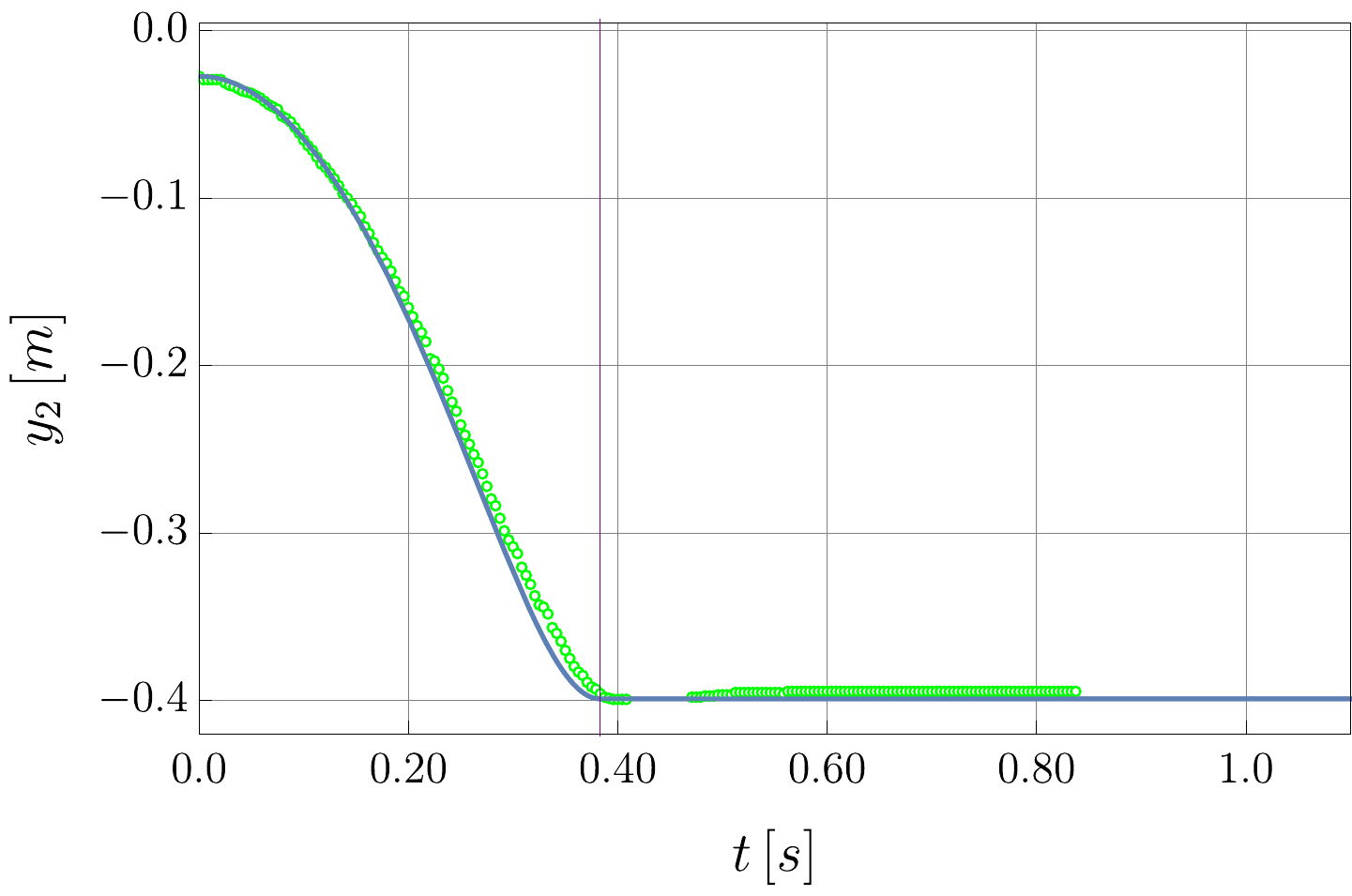}
         \caption{}
         \label{fig:M2Dynamics}
     \end{subfigure} \\
     \begin{subfigure}[b]{0.48\textwidth} 
         \centering
         \includegraphics[width=\textwidth]{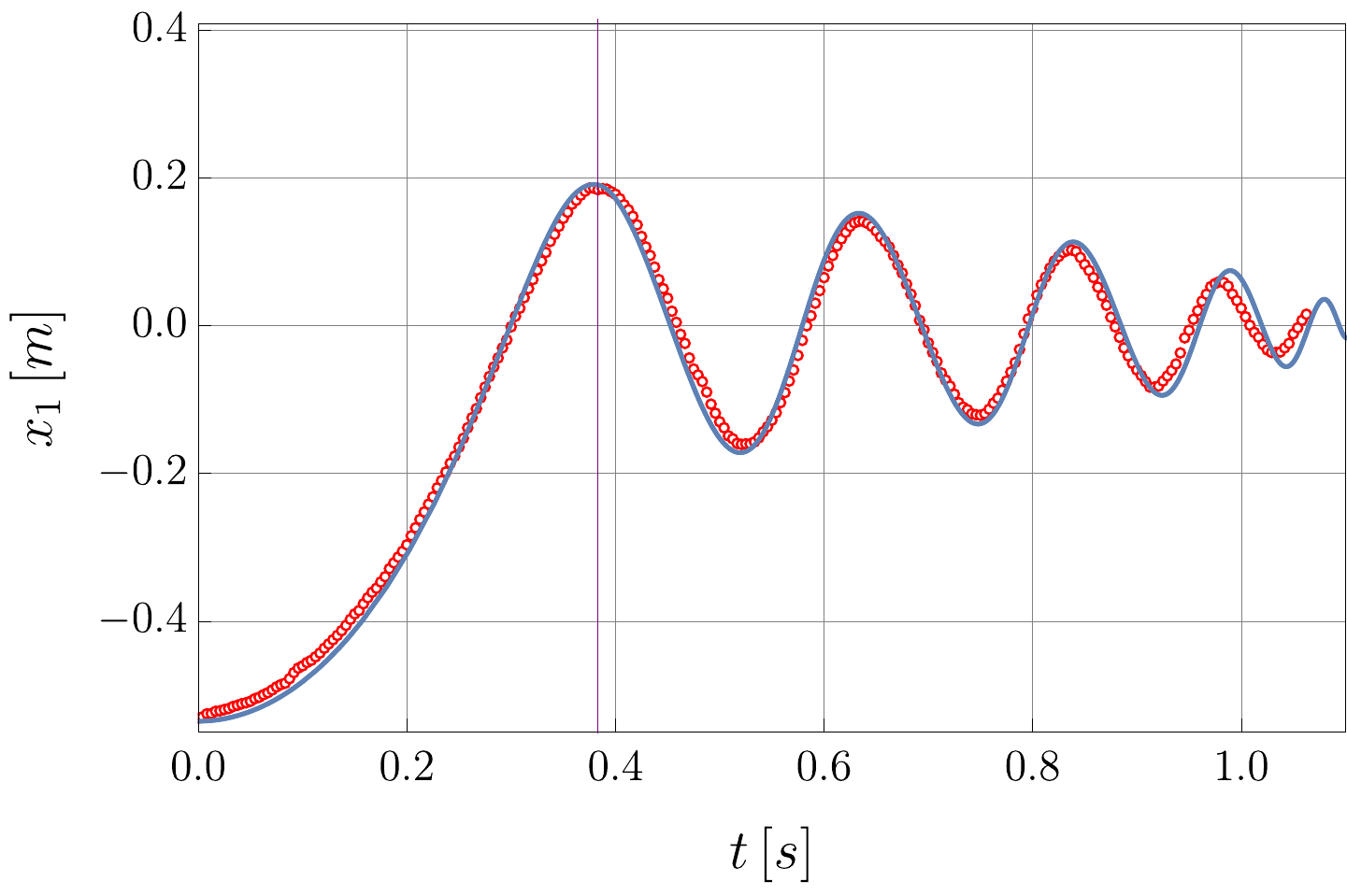}
         \caption{}
         \label{fig:x1Dynamics}
     \end{subfigure}
     \hfill
     \begin{subfigure}[b]{0.48\textwidth}
         \centering
         \includegraphics[width=\textwidth]{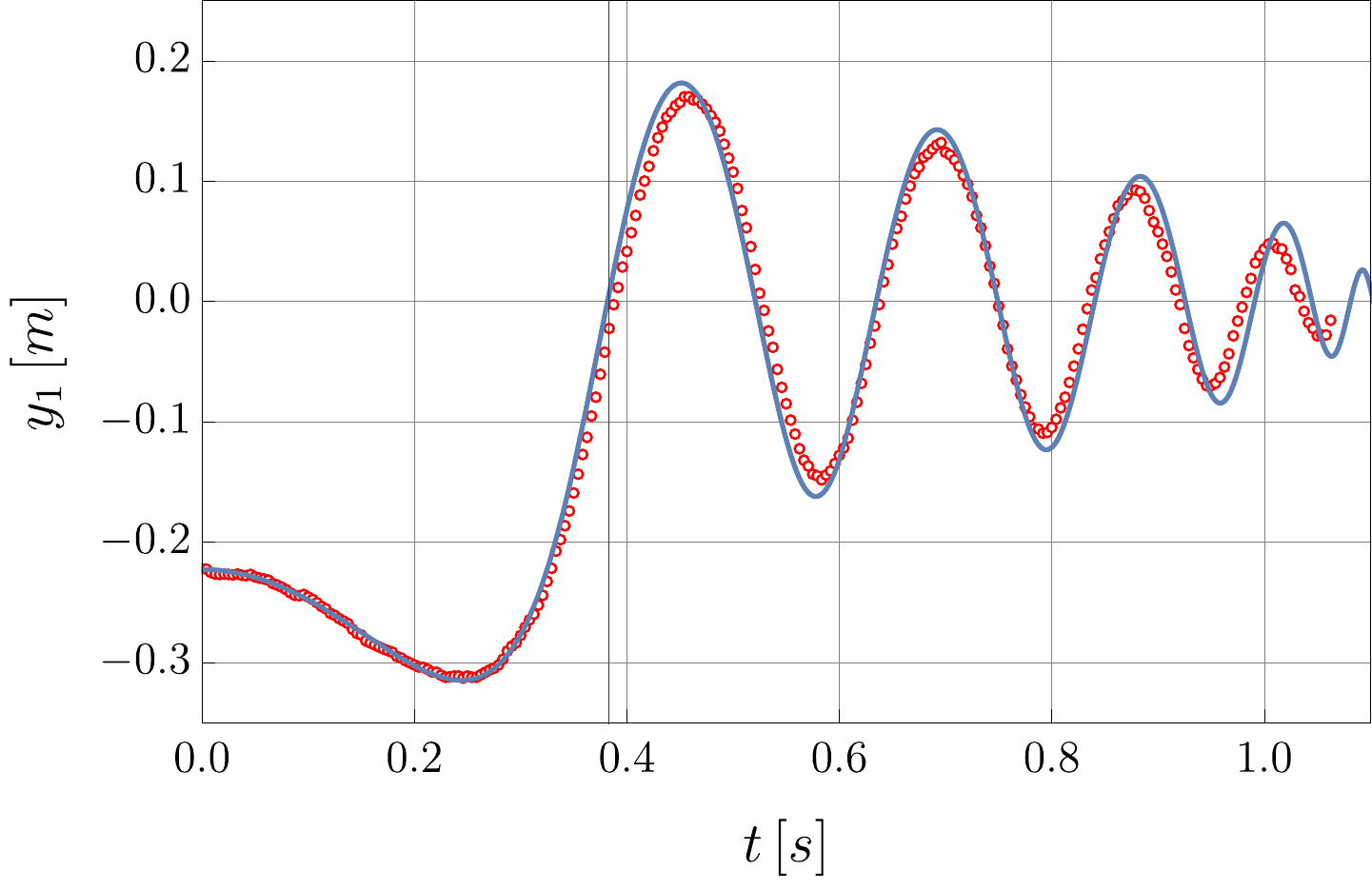}
         \caption{}
         \label{fig:y1Dynamics}
     \end{subfigure}
        \caption{These plots show experimental data for $M_1$ (red circles) and $M_2$ (green circles), along with the numerical solution of our model (blue curve), for $M_2 / M_1 = 10$. The vertical purple line in Figs.~\ref{fig:M2Dynamics}, \ref{fig:x1Dynamics}, and \ref{fig:y1Dynamics} indicates the time when $M_2$ halts.}
        \label{fig:10to1Comparison}
\end{figure}
Data for a typical trial with mass ratio $M_2 / M_1 = 10$ ($M_1 = 10\,\text{g}$, $M_2 = 100\,\text{g}$) is compared with the numerical solution of our model in Fig.~\ref{fig:10to1Comparison}. The first subfigure, Fig.~\ref{fig:M1Trajectory}, shows the path followed by $M_1$ as it loops around the rod. The simulation accurately tracks the measured position of $M_1$ from release until the point where $M_2$ halts (the solid black dot), and then precisely replicates the spiral produced by the string wrapping around the rod (see also Appendix \ref{app:SubsequentMotion}). The primary source of error affecting these plots is uncertainty in accurately judging the position of $M_1$, caused by the blurring of the mass in the video.

Figure \ref{fig:M2Dynamics} compares the observed and simulated dynamics of $M_2$ as it falls vertically. The model predicts that the mass drops a distance $\Delta y_{2,\ts{num}} = -0.37\,\text{m}$ before halting, and the observed value was $\Delta y_{2,\ts{exp}} = -0.37\,\text{m}$. The numerical solution predicts that $M_2$ halts $0.38\,\text{s}$ after $M_1$ is released, which is again in good agreement with the observed halt time of $0.40\,\text{s}$. However, in Fig.~\ref{fig:M2Dynamics} the mass can be seen to pass directly through its final $y_2$ value closer to the predicted halt time. It briefly dips below its final $y_2$ value and then returns to the same point at a slightly later time. We suspect that this behavior is caused by the slight elasticity of the string, which was not accounted for in the model. 

Finally, Figs.~\ref{fig:x1Dynamics} and \ref{fig:y1Dynamics} show the $x$ and $y$ positions of mass $M_1$ as a function of time. As with $M_2$, the numerical solution precisely captures the dynamics of $M_1$ as it loops around the rod. The numerical solution predicts that $M_1$ reaches the rod ($\ell=0$) at $t_\ts{end} = 1.1\,\text{s}$. This prediction can't be directly compared to experiment, since the finite size of $M_1$ in our apparatus causes it to collide with the rod before the length $\ell$ reaches zero. However, we can compare the times at which $x_1$ and $y_1$ reach local extrema where $v_{1,x} = 0$ or $v_{1,y} = 0$. These times are shown in Table \ref{M1Data} of Appendix \ref{app:Data} and the predictions of the numerical solution are in excellent agreement with the observed values. We should emphasize that the effects of air resistance are essential for the accuracy of the model. Without air resistance, the model's prediction for when $M_1$ reaches the rod is typically $5-10\%$ earlier than with air resistance included. Although this might amount to less than a tenth of a second, it is enough for the late time behavior of the simulation to move completely out of phase with the data. To faithfully track the dynamics of $M_1$, air resistance must be included in the model.

\begin{figure}[htb]
     \centering
     \begin{subfigure}[b]{0.48\textwidth}
         \centering
         \includegraphics[width=\textwidth]{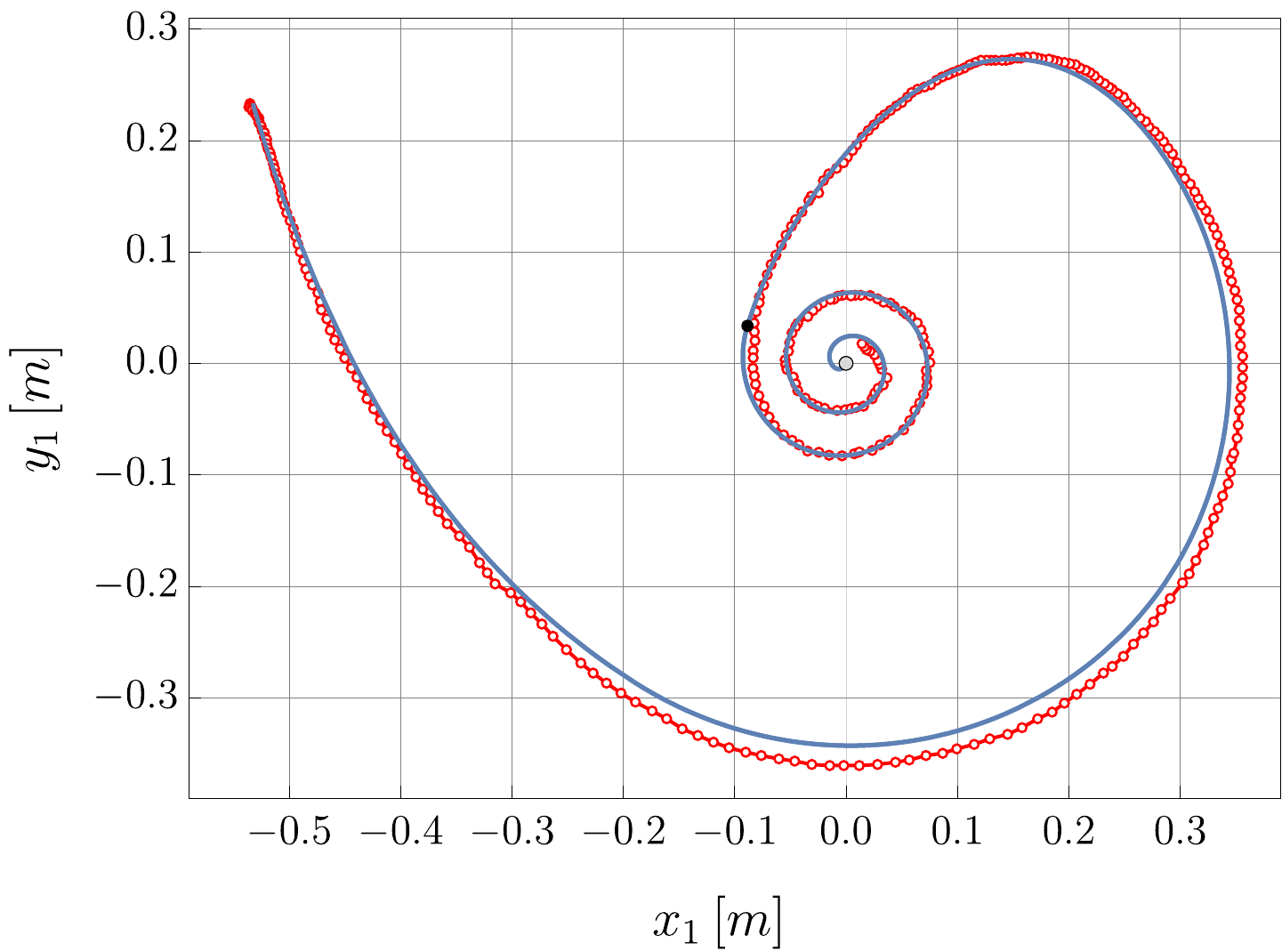} 
         \caption{}
         \label{fig:M1Trajectory2to1}
     \end{subfigure}
     \hfill
     \begin{subfigure}[b]{0.48\textwidth}
         \centering
         \includegraphics[width=\textwidth]{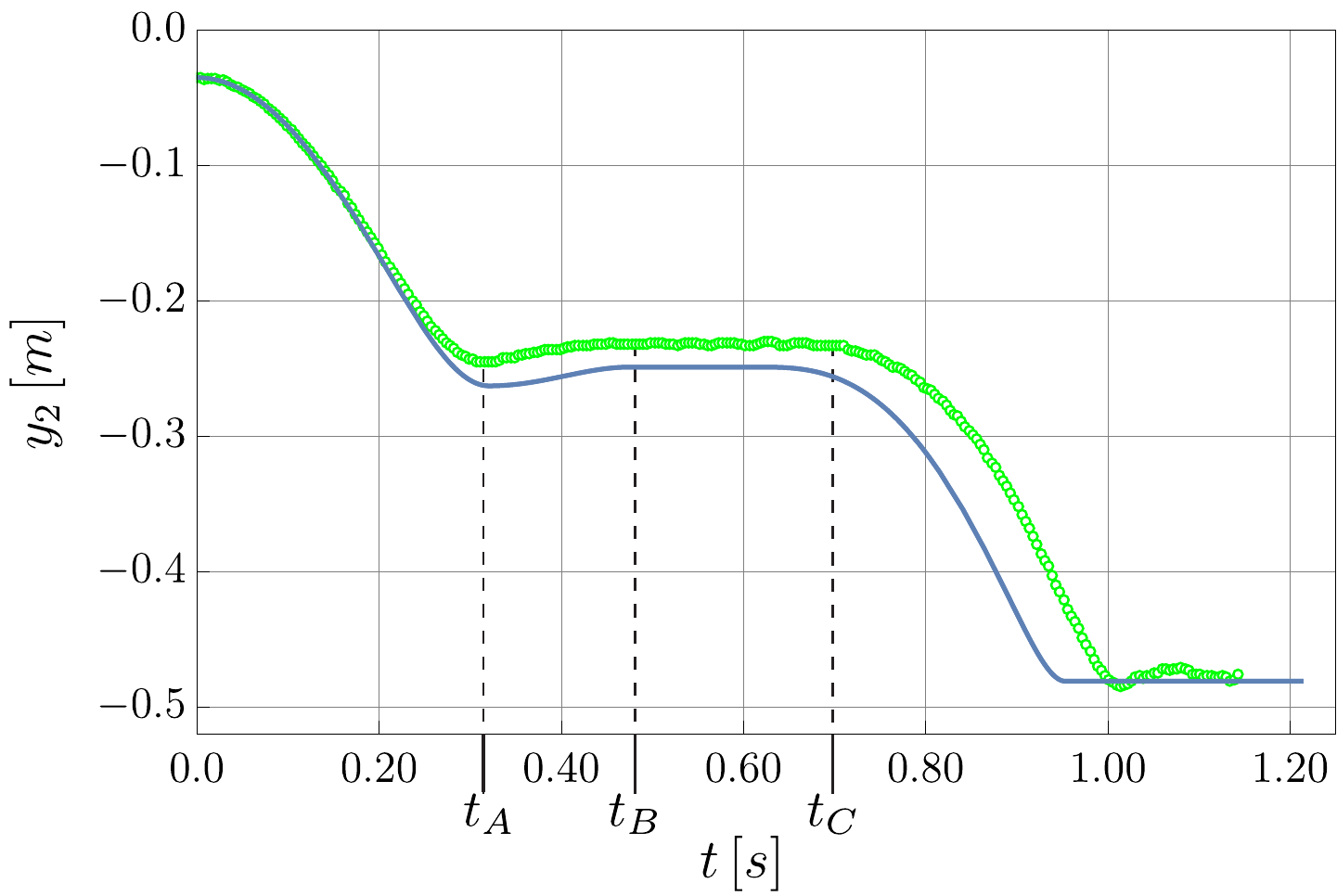}
         \caption{}
         \label{fig:M2Dynamics2to1}
     \end{subfigure} 
        \caption{\label{fig:2to1Comparison} The numerical solution captures nontrivial features of the trajectory and dynamics of the masses for the mass ratio $M_2 / M_1 = 2$.}
\end{figure}

Our experiments show that the behavior of the looping pendulum for large values of $M_2/M_1$ usually involves a quick halt, followed by a regular spiral like the one in Fig.~\ref{fig:M1Trajectory}.~\footnote{The shape traced out by $M_1$ after $M_2$ halts can also be understood in terms of the geometrical considerations discussed in Appendix \ref{app:SubsequentMotion}.} However, the looping pendulum exhibits more complicated behavior at smaller values of $M_2/M_1$, which is also well accounted for in our model.  

An example is shown in Fig.~\ref{fig:2to1Comparison} for the case $M_2 / M_1 = 2$ ($M_1 = 50\,\text{g}$, $M_2 = 100\,\text{g}$). The mass $M_1$ follows a path with a qualitatively different shape than the trajectory shown in Fig.~\ref{fig:M1Trajectory}. Nevertheless, it is well described by the numerical solution. The motion of $M_2$ is also very different than what is seen in Fig.~\ref{fig:M2Dynamics}. It is observed to drop, briefly stop at time $t_A$, then move upwards before stopping again at time $t_B$, and then resume falling at time $t_C$ before finally halting. The same behavior appears in the numerical solution of our model. Table \ref{M2Data} in Appendix \ref{app:Data} summarizes the observed and predicted values for the times and positions of these events. The dynamics match up well for the first few events, with small differences between the predicted and observed values for $t_A$ and $t_B$. There is a larger disagreement between the observed and predicted values for the time $t_C$ at which $M_2$ begins its final downward plunge. But the model's halting time is once again in good agreement with the data, and it successfully reproduces the final distance that $M_2$ drops. (There appears to be another small ``bounce'' in the data as $M_2$ abruptly stops, likely due to stretching of the string.) Despite slightly larger errors for some of the intermediate features in $M_2$'s trajectory, the model gives a compelling description of what was observed in the experiment. In fact, the $M_2/M_1 = 2$ trial was carried out because the simulation predicted this interesting behavior for $M_2$.

\section{Discussion}
\label{sec:Disc}

In this paper we developed a model of the looping pendulum using Newton's Laws. The result is a system of two coupled, nonlinear differential equations which we solved numerically. The numerical solution accurately describes the trajectory and dynamics of the masses, and reproduces complex behaviors observed for smaller values of the ratio $M_2/M_1$.

The apparatus described in Section \ref{sec:ExpSetup} was built with materials that were on hand. Replacing some of the components would reduce uncertainty in data collection, eliminate the effects of forces we did not account for, and improve our ability to estimate parameters that appear in the model. 

The single largest source of uncertainty when collecting position data was blurring of the fast-moving masses (especially $M_1$) in the video. Future versions of the experiment will rely on a dedicated high speed camera instead of an iPhone, which will reduce blurring and allow us to more accurately identify the positions of the masses in each frame. 

Another source of error is the slight elasticity of the string. The string connecting the masses will be replaced with a material that is less susceptible to stretching. We expect that this will reduce the small bounce seen in $M_2$'s position as it halts, resulting in more reliable measurements of the halting time. 

A final challenge is factoring in air resistance, which proved important for reproducing the dynamics of the pendulum. The orientations of the masses, and hence the area they present in a plane perpendicular to their velocity, change throughout their motion. This alters the drag coefficients, which are already difficult to estimate. Replacing bundles of washers with cylindrical masses was an improvement, but spherical masses that present the same area throughout their motion would better address this problem.\footnote{One could also conduct the experiment in a vacuum, but this seems impractical.}

Using an improved apparatus, we will explore a larger range of the looping pendulum's parameter space. These first investigations sampled a relatively small number of initial angles for mass ratios $2 \leq M_2 / M_1 \leq 10$. We hope to extend this to higher and lower values of $M_2/M_1$, and to explore a much larger set of initial angles. For lower values of $M_2/M_1$ the system only seems to halt for isolated values of the initial angle, so it would be interesting to see if there is a nontrivial value of $M_2/M_1$ below which halting does not occur at all. Also, the presence of air resistance in the model means that the individual masses are important, as opposed to just their ratio. It would also be interesting to explore the behavior of the system as $M_1$ and $M_2$ are increased (or decreased) with $M_2/M_1$ held fixed. The ratios $\gamma_1 / M_1$ and $\gamma_2 / M_2$ appear in the equations of motion, so the area scaling of $\gamma$ and volume scaling of $M$ should make air resistance less important for larger masses.

Finally, at smaller values of $M_2/M_1$ our model seems to show sensitivity to small changes in initial conditions. This is reminiscent of simple systems that display chaotic behavior, like the double-pendulum. For larger values of $M_2/M_1$ the numerical solutions are robust in the sense that small changes in the initial angle (or other parameters of the model) lead to small changes in the trajectory and quantities like $t_\ts{halt}$. But as we decrease $M_2/M_1$ these ``nearby'' solutions often describe qualitatively different behaviors. 
\begin{figure}[h!]
     \centering
     \begin{subfigure}[b]{0.49\textwidth}
         \centering
         \includegraphics[width=\textwidth]{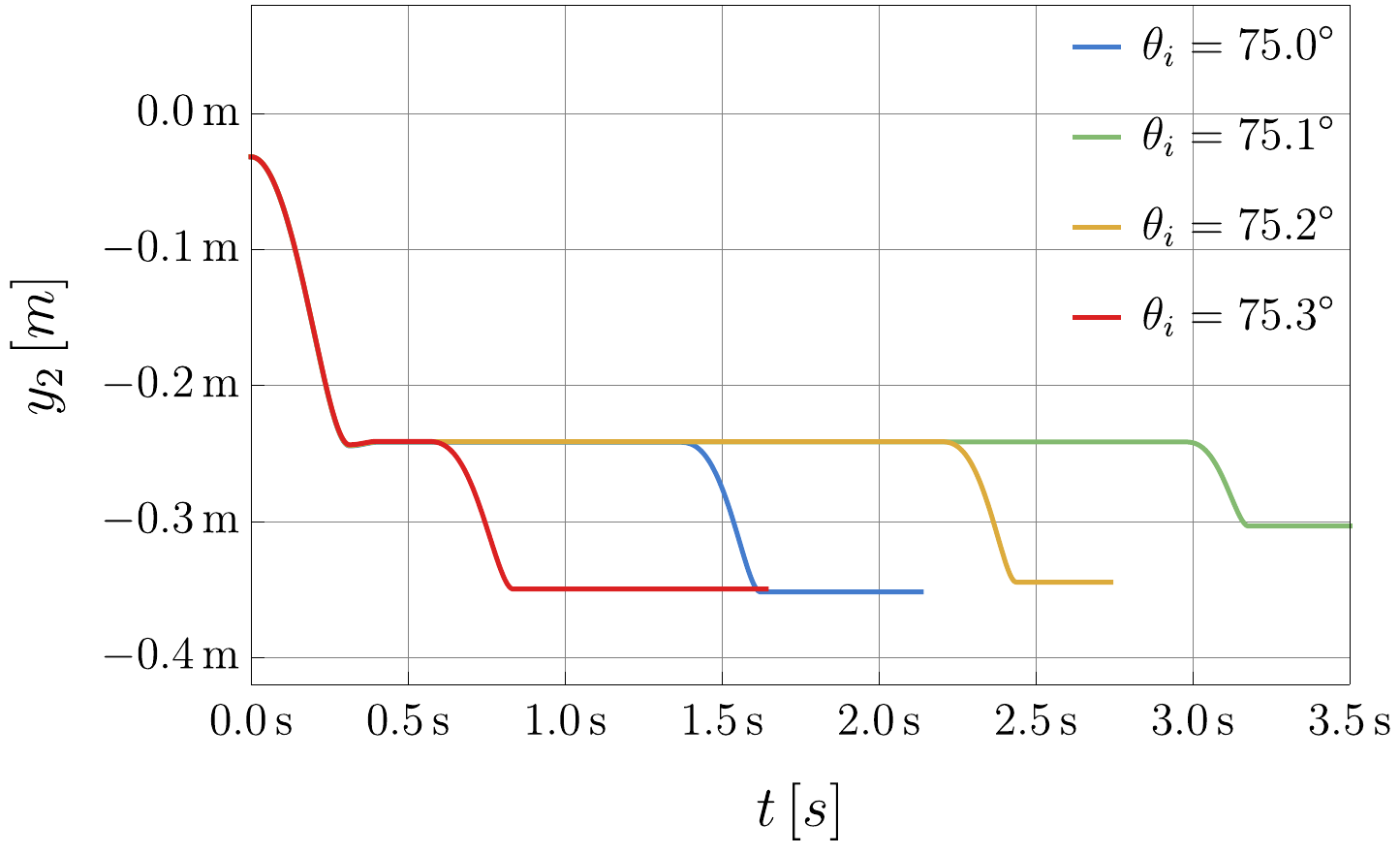} 
     \end{subfigure}
     \hfill
     \begin{subfigure}[b]{0.49\textwidth}
         \centering
         \includegraphics[width=\textwidth]{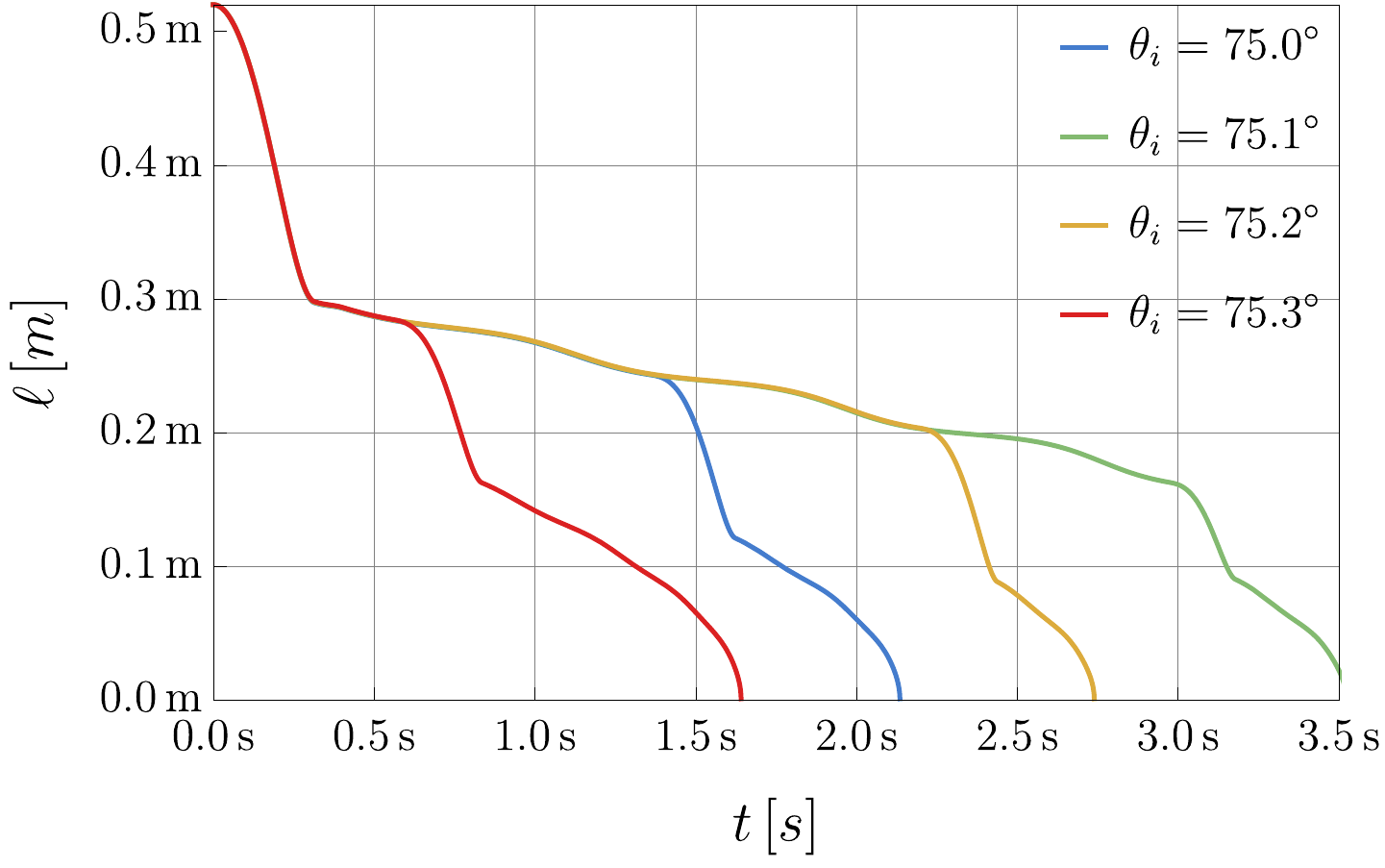}
     \end{subfigure} 
        \caption{The simulated behavior of $y_2$ and $\ell$ for $M_2/M_1=2.5$ and initial angles $75^{\circ}$, $75.1^{\circ}$, $75.2^{\circ}$, and $75.3^{\circ}$ suggests sensitive dependence on the initial conditions.}
		\label{fig:5to2Comparison}
\end{figure}
Figure \ref{fig:5to2Comparison} shows an example of this for $M_2 / M_1 = 2.5$ ($M_2 = 38\,\text{g}$, $M_1 = 15.2\,\text{g}$). Beginning at an initial angle of $75^{\circ}$ and incrementing in steps of $0.1^{\circ}$, the solutions exhibit the same initial behavior before branching out into a wide range of halting times. It is not clear yet whether this is a genuine feature of the looping pendulum or an artifact of the numerical methods used to solve the equations of motion. But we suspect that the looping pendulum may exhibit chaotic behavior, and we are currently investigating this possibility.

\begin{acknowledgments}

We wish to thank Loyola University Chicago for facilitating lab access during the COVID-19 pandemic shutdown, and the Department of Physics for providing the materials and work space used in the experiments.

\end{acknowledgments}

\appendix   
\section{Data}
\label{app:Data}

Table \ref{M1Data} shows the observed and predicted times for the local extrema of $x_1$ and $y_1$ in Figs.~ \ref{fig:x1Dynamics} and \ref{fig:y1Dynamics}, respectively. Times in the $t_\ts{exp}$ columns were found by identifying points with the largest local values (positive or negative) of $x_1$ and $y_1$ in the data. The values in the $t_\ts{num}$ columns were obtained by evaluating the derivatives $\dot{x}_{1}$ and $\dot{y}_{1}$ for the numerical solution and finding their roots. As explained in the text, the numerical solution is in excellent agreement with the data. Table \ref{M2Data} records the times and positions for various events experienced by mass $M_2$ as shown in Fig.~\ref{fig:M2Dynamics2to1}.
\begin{table}
	\begin{tabular}{|c|c|c|c|c|}
		\hline
		\, ~ \,  &
		\multicolumn{2}{c|}{$v_{1,x} = 0$} &
		\multicolumn{2}{c|}{$v_{1,y} = 0$} \\ \hline
		& \quad $t_\ts{exp}$ ~\,  & \quad $t_\ts{num}$ ~\,  & \quad $t_\ts{exp}$ ~\,  & \quad  $t_\ts{num}$ ~\,  \\
		\hline
		1 & $0.38\,\Ts$ & $0.38\,\Ts$ & $0.25\,\Ts$ & $0.24\,\Ts$  \\
		2 & $0.52\,\Ts$ & $0.52\,\Ts$ & $0.45\,\Ts$ & $0.45\,\Ts$ \\
		3 & $0.63\,\Ts$ & $0.63\,\Ts$ & $0.58\,\Ts$ & $0.58\,\Ts$  \\
		4 & $0.75\,\Ts$ & $0.75\,\Ts$ & $0.70\,\Ts$ & $0.69\,\Ts$  \\
		5 & $0.84\,\Ts$ & $0.84\,\Ts$ & $0.79\,\Ts$ & $0.79\,\Ts$  \\
		6 & $0.91\,\Ts$ & $0.92\,\Ts$ & $0.87\,\Ts$ & $0.88\,\Ts$  \\
		7 &  $0.98\,\Ts$ & $0.99\,\Ts$   & $0.95\,\Ts$  &  $0.96\,\Ts$  \\
		\hline
	\end{tabular}
	\caption{\label{M1Data} Experiment and numerical solution values for the times when $x_1$ and $y_1$ experience local extrema in Figs.~\ref{fig:x1Dynamics} and \ref{fig:y1Dynamics}.}
\end{table}

\begin{table}
	\begin{tabular}{|c|c|c|c|}
		\hline
		~ Quantity \, 				& \, Experiment \, 			& \, Numerical \, 		& \, Percent Error ~ \\ \hline
		$t_A$ 						& $0.31\,\text{s}$ 			& $0.32\,\text{s}$  	& $4.6$ \\
		$\Delta y_2 (t_A)$ 			& $-0.21\,\text{m}$ 		& $-0.23\,\text{m}$ 	& $8.5$ \\
		$t_B$ 						& $0.46\,\text{s}$	 		& $0.48\,\text{s}$ 		& $3.0$ \\
		$\Delta y_2 (t_B)$ 			& $-0.20\,\text{m}$ 		& $-0.21\,\text{m}$ 	& $8.6$ \\
		$t_C$ 						& $0.71\,\text{s}$  		& $0.62\,\text{s}$  	& $12.7$ 		\\
		$t_\ts{halt}$ 				& $1.0\,\text{s}$ 			& $0.95\,\text{s}$  	& $4.6$ \\
		$\Delta y_2 (t_\ts{halt})$ 	& $-0.44\,\text{m}$ 		& $-0.44\,\text{m}$ 	& $0.7$ \\
		\hline
	\end{tabular}
	\caption{\label{M2Data} Observed and predicted values for the times and changes in position at different points along the trajectory of $M_2$ as illustrated in Fig.~\ref{fig:M2Dynamics2to1}.}
\end{table}

\section{Motion of $M_1$ After Halting}
\label{app:SubsequentMotion}

\begin{figure}[h!]
  \centering
	\includegraphics[width=0.5\columnwidth]{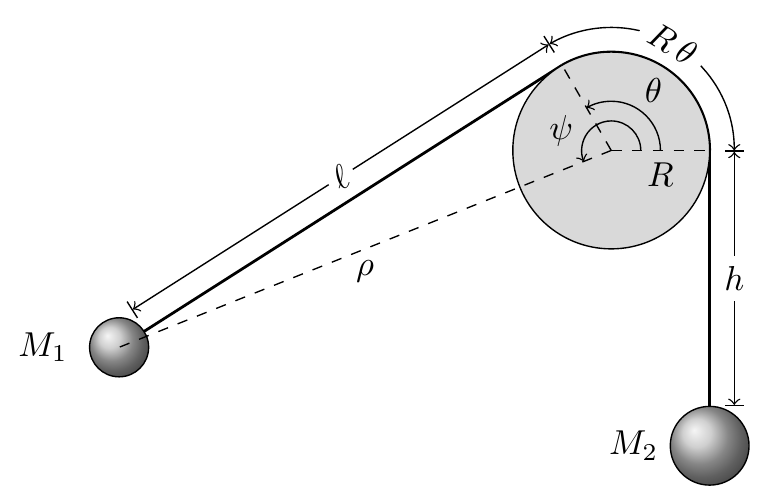}
	\caption{  \label{PendulumHaltedTrajectory} After $M_2$ has halted, the subsequent path of $M_1$ is determined by the fixed length of string $\ell_\ts{halt}$ looping around the rod. The polar coordinates $(\rho,\psi)$ shown here are used to describe this path.}
\end{figure}

\begin{figure}[h!]
	\centering
	\includegraphics[width=0.5\columnwidth]{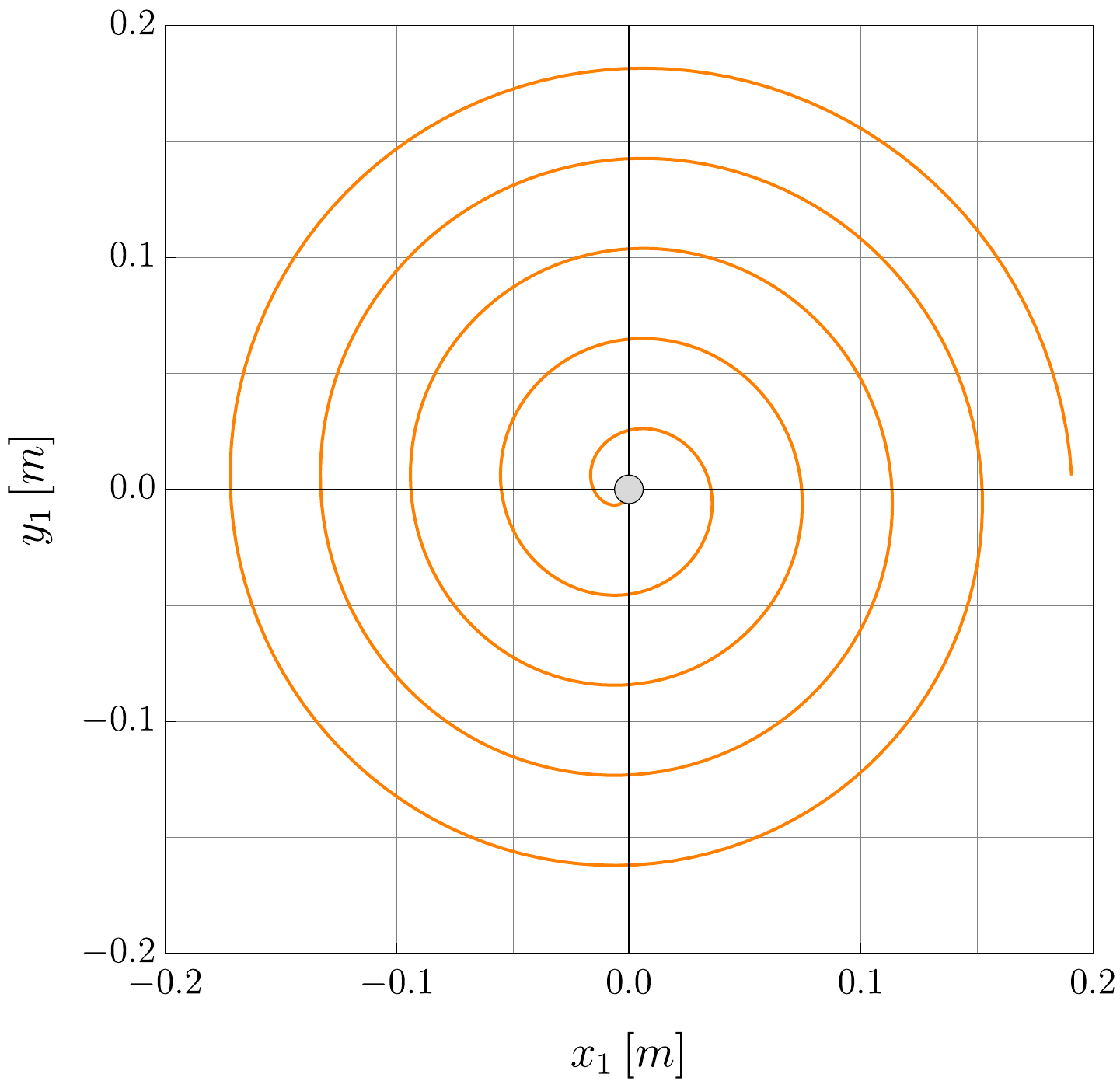}
	\caption{After mass $M_2$ halts in Fig.~\ref{fig:M1Trajectory}, mass $M_1$ follows the spiral described by Eqs.~\eqref{eq:HaltedM1Trajectory1}-\eqref{eq:HaltedM1Trajectory2} with $\ell_\ts{halt} = 0.196\,\text{m}$ and $\theta_\ts{halt} = 4.738$.}
	\label{HaltedSpiral} 
\end{figure}

Once the heavier mass halts, the shape of the path followed by the lighter mass is completely determined by the string wrapping around the rod. The shape of this spiral can be described using the polar coordinates $(\rho,\psi)$ shown in Fig.~\ref{PendulumHaltedTrajectory}, where $\rho = \sqrt{R^2 + \ell^2}$ is the distance from the center of the rod to $M_1$ and $\psi = \theta + \arctan(\ell/R)$ is its angular position.
Let $\ell_\ts{halt}$ and $\theta_\ts{halt}$ be the values of $\ell$ and $\theta$, respectively, when $M_2$ halts. Since $h$ is now constant, $L-h = \ell + R\,\theta = \ell_\ts{halt} + R\,\theta_\ts{halt}$ and the shape of the curve is
\begin{gather}\label{eq:HaltedM1Trajectory1}
	\rho(\psi) = \sqrt{R^2 + (\ell_\ts{halt} + R\,\theta_\ts{halt}-R\,\theta(\psi))^2} ~,
\end{gather} 
where $\theta(\psi)$ is defined implicitly by the transcendental equation
\begin{gather}\label{eq:HaltedM1Trajectory2}
	\psi = \theta(\psi) + \arctan\Big(\frac{\ell_\ts{halt}}{R} + \theta_\ts{halt} - \theta(\psi)\Big) ~.
\end{gather}
The value of $\theta$ when $M_1$ reaches the rod ($\ell=0$) is $\theta_\ts{halt} + \ell_\ts{halt}/R$, so $\psi$ takes values in the range
\begin{gather}
	\theta_\ts{halt} + \arctan\Big(\frac{\ell_\ts{halt}}{R}\Big) \leq \psi \leq \theta_\ts{halt} + \frac{\ell_\ts{halt}}{R} ~.
\end{gather}
The values $\ell_\ts{halt}$ and $\theta_\ts{halt}$ depend on the dynamics of the system, and for some initial conditions halting may not occur at all. But once $M_2$ has completely stopped moving, the shape of the curve traced out by $M_1$ is geometrically determined by the string looping around the rod. As an example, Fig.~\ref{HaltedSpiral} reproduces this path for $M_1$ in the $M_2/M_1 = 10$ trial shown in Fig.~\ref{fig:M1Trajectory}.

Once $M_2$ has completely halted we can also give an approximate solution for $\ell(t)$ and $\theta(t)$ in the case where air resistance is negligible. As above, $\ell + R\,\theta = \ell_\ts{halt} + R\,\theta_\ts{halt}$ is constant and therefore $\dot{\ell} + R\,\dot{\theta} = \ddot{\ell} + R\,\ddot{\theta} = 0$.
Then the $\hat{t}_1$ component of the equation of motion for $M_1$ becomes
\begin{gather} 
	\dot{\ell}\,\dot{\theta} + \ell\,\ddot{\theta} = g\,\sin \theta ~,
\end{gather}
which can be rewritten as
\begin{gather}\label{HaltedEOM1}
	\dot{\ell}^{2} + \ell\,\ddot{\ell} = - g\,R\,\sin\left(\theta_\ts{halt} + \frac{\ell_\ts{halt}}{R} - \frac{\ell}{R}\right) ~.
\end{gather}
If we ignore the right-hand side of the equation then the solution from the time $t_\text{\tiny halt}$ when $M_2$ halts until the time $t_\text{\tiny end}$ when $M_1$ reaches the rod is
\begin{gather} \label{l0}
	\ell(t) = \ell_\ts{halt}\,\sqrt{\frac{t_\text{\tiny end} - t}{t_\text{\tiny end} - t_\text{\tiny halt}}} \\ \label{theta0}
	\theta(t) = \theta_\ts{halt} + \frac{\ell_\ts{halt}}{R} - \frac{\ell_\ts{halt}}{R}\,\sqrt{\frac{t_\text{\tiny end} - t}{t_\text{\tiny end} - t_\text{\tiny halt}}}~.
\end{gather}
For the cases we have considered these expressions differ from the numerical solution of the system by no more than a few percent. This suggests that an analytic solution of the full equation Eq.~\eqref{HaltedEOM1} can be obtained perturbatively. However, our experiments show that the effects of air resistance are important for capturing the dynamics of the system, so we will not pursue this further.


\begin{thebibliography}{4}%
\makeatletter
\providecommand \@ifxundefined [1]{%
 \@ifx{#1\undefined}
}%
\providecommand \@ifnum [1]{%
 \ifnum #1\expandafter \@firstoftwo
 \else \expandafter \@secondoftwo
 \fi
}%
\providecommand \@ifx [1]{%
 \ifx #1\expandafter \@firstoftwo
 \else \expandafter \@secondoftwo
 \fi
}%
\providecommand \natexlab [1]{#1}%
\providecommand \enquote  [1]{``#1''}%
\providecommand \bibnamefont  [1]{#1}%
\providecommand \bibfnamefont [1]{#1}%
\providecommand \citenamefont [1]{#1}%
\providecommand \href@noop [0]{\@secondoftwo}%
\providecommand \href [0]{\begingroup \@sanitize@url \@href}%
\providecommand \@href[1]{\@@startlink{#1}\@@href}%
\providecommand \@@href[1]{\endgroup#1\@@endlink}%
\providecommand \@sanitize@url [0]{\catcode `\\12\catcode `\$12\catcode
  `\&12\catcode `\#12\catcode `\^12\catcode `\_12\catcode `\%12\relax}%
\providecommand \@@startlink[1]{}%
\providecommand \@@endlink[0]{}%
\providecommand \url  [0]{\begingroup\@sanitize@url \@url }%
\providecommand \@url [1]{\endgroup\@href {#1}{\urlprefix }}%
\providecommand \urlprefix  [0]{URL }%
\providecommand \Eprint [0]{\href }%
\providecommand \doibase [0]{http://dx.doi.org/}%
\providecommand \selectlanguage [0]{\@gobble}%
\providecommand \bibinfo  [0]{\@secondoftwo}%
\providecommand \bibfield  [0]{\@secondoftwo}%
\providecommand \translation [1]{[#1]}%
\providecommand \BibitemOpen [0]{}%
\providecommand \bibitemStop [0]{}%
\providecommand \bibitemNoStop [0]{.\EOS\space}%
\providecommand \EOS [0]{\spacefactor3000\relax}%
\providecommand \BibitemShut  [1]{\csname bibitem#1\endcsname}%
\let\auto@bib@innerbib\@empty
\bibitem [{Note1()}]{Note1}%
  \BibitemOpen
  \bibinfo {note} {The model presented in this paper was developed before the
  publication of reference \cite {Yubo}.}\BibitemShut {Stop}%
\bibitem [{Note2()}]{Note2}%
  \BibitemOpen
  \bibinfo {note} {Fixed values within small variations associated with
  positioning the string, attaching the masses, and using different
  rods}\BibitemShut {NoStop}%
\bibitem [{Note3()}]{Note3}%
  \BibitemOpen
  \bibinfo {note} {The shape traced out by $M_1$ after $M_2$ halts can also be
  understood in terms of the geometrical considerations discussed in Appendix
  \ref {app:SubsequentMotion}.}\BibitemShut {Stop}%
\bibitem [{Note4()}]{Note4}%
  \BibitemOpen
  \bibinfo {note} {One could also conduct the experiment in a vacuum, but this
  seems impractical.}\BibitemShut {Stop}%
\end{thebibliography}%


\begin{thebibliography}{99}

\bibitem{YouTube} See for example the YouTube channel ``Homemade Science
	with Bruce Yany'' \url{https://youtu.be/SXQ9VaYm3yQ} which gathered over 1.4 M
	views since 2015.

\bibitem{IYPT} ``Looping Pendulum'' \url{https://www.iypt.org/problems/problems-for-the-32nd-iypt-2019}, problem 14.

\bibitem{Yubo} Zhou Yu-bo et al, ``Research on the looping pendulum phenomenon,'' European Journal of Physics \textbf{41} 025003 (2020). DOI: \url{https://doi.org/10.1088/1361-6404/ab5e68}

\bibitem{Mathematica} {Wolfram Research{,} Inc.}, Computer Program ``Mathematica,'' {V}ersion 12.2 (2020), \url{https://www.wolfram.com/mathematica}

\bibitem{Tracker} D.~Brown, Computer Program ``Tracker Video Analysis and Modeling Tool,'' Version 5.1.4 (2020), WWW Document, \url{https://physlets.org/tracker/}.

\end{thebibliography}
\end{document}